# Generative AI in Education: Student Skills and Lecturer Roles

Stefanie Krause[1][000-0002-1271-7514], a, Ashish Dalvi[1, b] and Syed Khubaib Zaidi[1, c]

[1] Harz University of Applied Sciences, 38855 Wernigerode, Germany
[a]skrause@hs-harz.de; [b]u39011@hs-harz.de; [b]u38122@hs-harz.de

**Abstract.** Generative Artificial Intelligence (GenAI) tools such as ChatGPT are emerging as a revolutionary tool in education that brings both positive aspects and challenges for educators and students, reshaping how learning and teaching are approached. This study aims to identify and evaluate the key competencies students need to effectively engage with GenAI in education and to provide strategies for lecturers to integrate GenAI into teaching practices. The study applied a mixed method approach with a combination of a literature review and a quantitative survey involving 130 students from South Asia and Europe to obtain its findings. The literature review identified 14 essential student skills for GenAI engagement, with AI literacy, critical thinking, and ethical AI practices emerging as the most critical. The student survey revealed gaps in prompt engineering, bias awareness, and AI output management. In our study of lecturer strategies, we identified six key areas, with GenAI Integration and Curriculum Design being the most emphasised. Our findings highlight the importance of incorporating GenAI into education. While literature prioritized ethics and policy development, students favour hands-on, project-based learning and practical AI applications. To foster inclusive and responsible GenAI adoption, institutions should ensure equitable access to GenAI tools, establish clear academic integrity policies, and advocate for global GenAI research initiatives.

## 1      Introduction

GenAI has emerged as a revolutionary technology with a wide array of applications, notably within the educational sector [3]. The advancement in technology has led to an increasing trend of change in educational practices, which is persistently transforming conventional teaching and learning approaches. Large Language Models (LLMs) represent artificial intelligence frameworks engineered to process, comprehend, and generate text that closely resembles human language [9]. These models are classified within a subset of deep learning paradigms that exploit neural network architectures, most notably transformers, to process extensive text data for multiple language-based operations such as translation, summarization, information retrieval and conversational systems. In contrast to conventional language models (LMs) that depend on task-specific supervised learning methodologies, pre-trained Language Models (PLMs) form the basis of LLMs because they use self-supervised learning on enormous datasets to establish universal language processing competencies before evolving into LLMs [28]. The primary characteristic of LLMs involves their enormous scale because their parameter counts exceed tens and hundreds of billions, while their training datasets reach multiple terabytes. These extensive capabilities empower LLMs to attain elevated levels of fluency, coherence, and contextual sensitivity, frequently nearing human-level efficacy in a variety of Natural Language Processing (NLP) tasks [28]. The development of LLMs receives support from three core elements, including transformer-based architecture development and improved processing capacity and expanded access to extensive training datasets. Cutting-edge models, such as GPT-4o [22], GPT-4o mini [22], Deepseek-R1 [33] and LLaMA 3 [33], serve as exemplary representations of the potency of LLMs in generating meaningful and contextually pertinent text, thereby positioning them as essential components in AI-driven applications across a multitude of fields.

The increasing adoption of GenAI technologies in education has raised concerns about their potential impact on learning outcomes and the utilization of these technologies [3]. Consequently, some educational institutions have restricted the use of AI platforms like ChatGPT in examinations and assessments to preserve teaching methods and student skills. For example, 43 university members of the 2022 Quacquarelli Symonds (QS) World University Rankings have restricted access to ChatGPT unless they allow its use for certain tasks [37]. Nevertheless, a relatively beneficial approach is to help students develop responsible and ethical approaches to applying GenAI in their learning process in accordance with the updated educational requirements [9].

The capabilities of GenAI have already been tested in different scenarios that include scoring well in competitive academic assessments, including the Wharton Business School examination, where it obtained a B to B+ Grade [38]. Such achievements emphasize the need for educators and higher education policymakers to rethink conventional teaching approaches and consider how AI tools can be utilised to enhance student learning experiences. Some innovative strategies could be such as educating students on responsible AI practices, incorporating projects-based tasks in the assessments and leveraging AI to evaluate academic performance. Nevertheless, there are some challenges, such as biases in the data on which GenAI models are trained, ChatGPTs creating hallucinations that lead to inaccuracies in outputs, which causes misleading insights and somehow wrong decision-making that would affect student learning negatively [3].

Although GenAI has numerous applications across the educational sector, from elementary schools to universities, there is a noticeable gap in comprehensive research addressing two critical areas: student skills necessary to interact with GenAI and educator's strategies to incorporate GenAI into learning so this study aims to address these challenges by identifying the essential skills students need to work effectively with GenAI, assessing their current competency skill level. Additionally, we examine how educators can adapt their teaching strategies to integrate GenAI technologies responsibly into their educational practices, overcoming barriers and maximizing their educational potential. This paper also provides specific suggestions to lec-turers to fill the gap in the use and application of GenAI in learning. These recommendations were developed based on a categorization of the strategies most commonly found in the literature, classified in such a way as to facilitate their easy application in various educational settings. In achieving these objectives, the research aims to close the existing gaps by addressing the aspects and subsequently supporting the implementation of GenAI for learning environments. The **main contributions** of our paper are:

- evaluating the **key competencies students need** to effectively engage with GenAI in education
- Providing **strategies for lecturers** to integrate GenAI into teaching practices.

The remainder of the paper is structured as follows: Section 2 highlights the background and related work along with the research questions, followed by Section 3, which covers a detailed description of our methodology. In Section 4, we present the results and analysis of our survey. Then in Section 5, we discuss our results and provide recommendations, in Section 6, we draw together the key conclusions.

## 2  Background and Related Work

GenAI is one of the most rapidly advancing subdomains of AI and Machine learning (ML), and it can revolutionize education. In tackling traditional problems of educative systems, GenAI offers strategic and practical approaches to improve the practice of teaching and learning. Recently, numerous large language models have been developed which have achieved very high levels of performance, like BERT [17], RoBERTa [18], LaMDA [19], BART [20], LLAMA [35] and DEEPSEEK [33]. Yet the exponential rate at which ChatGPT is growing has changed and influenced routine practices, particularly in the education sector. However, a key question that comes up in this process is how good the GenAI models are and how they are prone to incorporate biases, unfairness or subjectiveness. Unfortunately, the education sector is facing challenges at the moment because of the growth of new advanced technologies like GenAI. The development of GenAI has led to the creation of new innovative models, including the recently released ChatGPT, BART, Gemini (formerly known as Bard), and DeepSeek [16,33]. While GenAI offer innovative new solutions, it also raises ethical and academic integrity challenges, especially in the educational sector. Students may exploit AI systems for academic dishonesty, like creating complete essays, answering exam questions, or composing research content that may not easily be detected [13]. Considering the significant influence of these technologies on education sectors, particularly their capacity to generate human-like responses in specialized areas, students will require certain foundational competency skills, and instructors have to re-imagine the methods of teaching.

A wide range of prior research focuses on the impact of GenAI in education. For example, [1] discusses how ChatGPT is used for automating formative essay assessment and research support. Similarly, the authors in [23] explain how ChatGPT can be applied to support students in solving complex programming problems. However, researchers in studies [24] and [25] emphasize the risks of unverified AI outputs, underscoring the importance of expert validation before applying AI-generated content. Krause et al. [3], describe the potential positive and negative impacts of using GenAI to support the learning and engagement of students. According to their survey of 130 students, 87% thought that GenAI could contribute to a reduction in workload in achieving academic objectives; however, they had concerns about accuracy and cheating [3]. A master's-level course integrating GenAI significantly improved students' comfort and ethical understanding, with 100% of participants feeling comfortable post-course compared to 37.5% pre-course [5]. The AI-ICE Framework (Ideas, Connections, Extensions) used in the study revealed that the majority of the students worked within the "Ideas" category learning primary concepts of AI while a few Moved to the "Connections" level incorporating AI into their learning processes. Platforms like Packback demonstrate how GenAI fosters student engagement through inquiry-based learning [5].

The authors in [4] synthesized research on the advantages of GenAI, highlighting benefits such as personalized tutoring and automated grading. However, challenges such as biased outputs, privacy concerns, and limitations in training data remain significant barriers. Natural language processing models have advanced to the point where AI-generated text detectors have become ineffective, which forces educational institutions to modify their policies on responsible AI utilization. The current assessment methods will likely become insufficient to evaluate student skills, yet future research should explore integrating GenAI into teacher training and rethinking traditional assessments to maximize its potential [4]. Despite these challenges, researchers advocate for the responsible adoption of GenAI, complemented by human oversight and robust ethical guidelines, to enhance teaching and learning experiences [1]. Similarly, the study [7] demonstrated that generative conversational agents, like chatbots, improve learning outcomes by providing tailored feedback. A study in [2] refers to GenAI as an educational innovation within the student's autonomy, focusing on the role of GenAI in enhancing the study process of higher education institutions. By enabling students to personalize their educational journeys and actively engage in knowledge construction, ChatGPT fosters student-centered learning. Additionally, the study also emphasizes a lecturer strategy, fine-tuning AI for domain-specific applications for better outputs, and the need for students from all academic backgrounds to possess AI literacy skills to appropriately interact with GenAI tools such as ChatGPT. This includes understanding AI mechanisms, critically evaluating outputs, and addressing ethical, social, and practical implications. Self-regulated learning serves as a necessary mechanism to guarantee AI supports independent problem-solving rather than replacing it to attain responsible AI-integrated education [2].

The research in [8] suggests that institutions should regulate the use of GenAI through policy to allow its use for creative work, including writing, while maintaining academic integrity. It points out the necessity of guiding students and using GenAI to enhance their critical thinking abilities, as well as for students interacting with AI-generated outputs. The study [9] provides a comprehensive evaluation of GenAI in education and claims that GenAI is poised to transform learning through intelligent systems. Some of the real-life use cases are Duolingo, where GenAI is used for language teaching, and Quillionz, where GenAI is used to generate quizzes and create educational content. Similarly, the research [7] presents that responsive chatbots functioning as teaching assistants and companions could reduce student isolation more effectively than traditional teacher counseling. Moreover, AI tools trained on specific course materials, such as ProfBot and MathGPT, show promising potential in personalized learning. These domain-specific GenAI applications act as study companions, offering tailored explanations and interactive learning experiences [14]. Similarly, strategy like testing GenAI efficiency in educational practices is already been implemented in an AI-driven learning platform at a Chinese university, which utilized machine learning to analyze student behavior and performance, generating personalized learning materials and adaptive pathways that adjusted dynamically to individual progress [6].

In [8], the author investigates the transformative potential of GenAI within the realm of higher education, highlighting both prospects and obstacles. The authors highlight how GenAI, encompassing platforms such as ChatGPT, is becoming as ubiquitous as the internet, necessitating innovative approaches to teaching, learning, and assessment. However, the study presents a use case that implicit biases encoded in training data negatively affect mainly minorities, and that includes international scholars and those whose first language is not English. In addition, the effectiveness of existing AI detection tools, including Turnitin and OpenAI, is questionable due to high levels of false identification that might lead to the discrimination of vulnerable learners [8]. One of the similar cases can be exemplified at Texas A&M University, where inaccurate AI-based plagiarism detection led to the wrongful failure of an entire class. Lecturers need to embrace responsible AI approaches to create fair and transparent assessment systems in their educational institutions [7]. The study [10] discusses possibilities of using ChatGPT in education and emphasizing the strategies of learning, and also educators to use GenAI in lesson planning while applying critical thinking in the assessment and promoting educational equity. By encouraging teachers to assess the AI-generated content for teaching and learning and applying critical reflective judgment, GenAI fosters personalized educational experiences for individual students. However, factors such as biases, errors or plagiarism risk require some clear guidelines as to how to handle them to ensure academic integrity. While GenAI tools such as ChatGPT promote equity through open educational resources.

While previous research has explored the impact of GenAI on education, there is a significant gap in understanding the specific skills students need to effectively engage with GenAI technologies and also effective strategies lecturers must adopt for successful integration of GenAI in education so our study specifically focuses on assessing the current level of essential skills students possess for effectively engaging with GenAI and determining the level of support they need to develop these skills further. Moreover, we explore strategies that lecturers can adopt to integrate GenAI effectively into educational practices, addressing both pedagogical and ethical considerations. This dual focus on student skill development and lecturer strategies will provide a comprehensive understanding of how GenAI can be responsibly and effectively leveraged in education.

## 2.1 Research Questions

Our study aims to explore how GenAI can be integrated into educational practices, focusing on its impact on students and the need for the adaption of teaching methods. The first research question (RQ1) investigates the essential skills students need to effectively engage with GenAI in their learning processes. Our research specifically focuses on assessing the current level of essential skills students possess for effectively engaging with GenAI and determining the level of support they need to develop these skills further. Building on this, the second research question (RQ2) examines how educators can update their teaching methods to integrate GenAI with student learning and support student learning in a rapidly evolving technological landscape. This research aims to show educators and policymakers how to use GenAI technology responsibly and effectively in educational environments. Accordingly, the research questions formulated are as follows:

- *RQ1*. What essential skills do students need to effectively engage with GenAI for educational purposes?
- *RQ2*. How can lecturers adapt their teaching strategies to integrate GenAI and support student learning effectively?

# 3 Methodology

This study adopts a mixed-methods approach, combining a literature review and a quantitative survey to explore essential student skills and lecturer strategies for integrating GenAI into education. Initially, 25 relevant papers were identified. After abstract reading, 15 articles were selected for in-depth analysis, identifying key student competencies and instructional methodologies. A frequency analysis determined the most discussed student skills and lecturer strategies. The top 10 student skills formed the basis of our survey design. The structured Google Form survey, consisting of 3 sections and using a 5-point Likert scale, collected data on students' background, self-assessed proficiency in these skills, and support needs for skill development. Data collection ensured participant anonymity. The survey findings were analyzed to identify competency gaps and areas requiring further support. These insights informed recommendations for enhancing student skills and guiding lecturers in adapting their teaching strategies to effectively integrate GenAI into educational practices.

# 4 Results and Analysis

## 4.1 Literature review findings

This subsection presents the outcome of the literature review, highlighting key themes and frequency analysis results. The literature review focused on two topics, lecturer strategies and essential student skills to effectively engage with GenAI.

**Lecturer Strategies.** The comprehensive literature review revealed a multitude of pedagogical strategies employed by educators for the incorporation of GenAI within educational contexts. Six categories of strategies were developed. Each category in this study encompasses distinct strategies, systematically ranked according to their prevalence, thereby indicating their significance within the analyzed literature. The overall distribution of these categories is illustrated in Fig. 1.

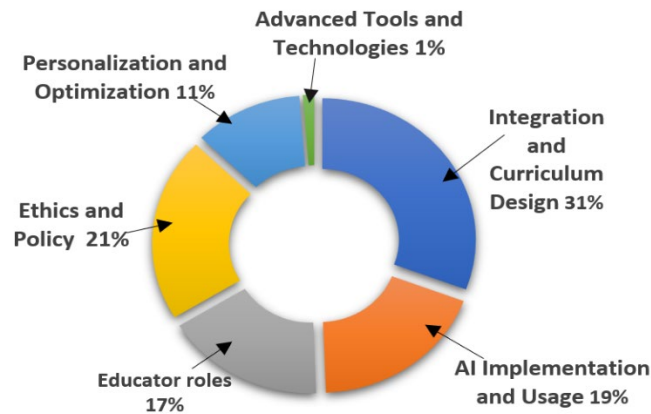

**Fig. 1.** Overall distribution of lecturer strategies by categories

**List of Strategies per Category.** The following tables (Tables 1-6) present a detailed breakdown of these strategies along with their respective frequencies, offering insight into the most commonly emphasized approaches for GenAI integration in educational settings.

**Table 1.** Strategies in Curriculum Design.

| Integration and Curriculum Design | Count |
|---|---|
| Integration into Curriculum | 13 |
| Adjust Exam and Assessment Strategies | 12 |
| Design Project-Oriented Tasks | 9 |
| Evaluating GenAI's Role in Pedagogical Practices | 7 |
| Introduce GenAI Gradually | 3 |
| Building Institutional Capacity for GenAI | 3 |
| Addressing Limitation of Resources | 3 |
| Framework for GenAI Implementation | 1 |
| **Grand Total** | **51** |

**Table 2.** Strategies in AI Implementation.

| AI Implementation and Usage | Count |
|---|---|
| AI-driven Feedback and Automated Grading | 8 |
| Testing the Efficiency of GPT in Educational Activities | 8 |
| Automating Routine Tasks | 6 |
| Using Chatbots for Learning Support | 5 |
| Implementing AI Tools for Interactive Classroom Learning | 4 |
| **Grand Total** | **31** |

**Table 3.** Strategies in Educator Roles

| Educator roles | Count |
|---|---|
| Teach AI Limitations | 9 |
| Upskilling for AI Proficiency | 7 |
| Rethink Learning Objectives | 6 |
| Serve as Mentors and Facilitators | 4 |
| Encourage Student Feedback on AI Experiences | 2 |
| **Grand Total** | **28** |

**Table 4.** Strategies in Ethics and Policy

| Ethics and Policy | Count |
|---|---|
| Policy Development for AI usage | 11 |
| Incorporate Ethics and Privacy in Curriculum | 10 |
| Integrate GenAI Responsibly | 7 |
| Enhancing Academic Integrity Tools for AI-Era Challenges | 4 |
| Implementing Explainable AI for Transparent and Controlled Outputs | 3 |
| **Grand Total** | **35** |

**Table 5.** Strategies in Personalization

| Personalization and Optimization | Count |
|---|---|
| Tailoring Pedagogical Strategies | 9 |
| Using Progress Reports to Optimize AI Tool Performance | 3 |
| Customizing AI Models with Subject-Specific Training | 3 |
| Personalized Quizzes Tailored to Student Performance | 3 |
| Creating Prompts for Questions Aligned with Learning Goals | 1 |
| **Grand Total** | **19** |

Table 6. Strategies in Advanced Tools

| Advanced Tools and Technologies | Count |
|---|---|
| Utilizing AI for Immersive Historical Content Creation | 1 |
| Harnessing AR, VR, and AI for Immersive Educational Experiences | 1 |
| **Grand Total** | **2** |

Overall, *the Integration and Curriculum Design* category was the most prominent since it appeared 51 times across 8 strategies, underscoring the emphasis on incorporating GenAI technologies into educational frameworks and aligning assessments. The importance of responsible GenAI implementation through practical usage strategies along with *ethical and policy* considerations emerges across 35 occurrences with 5 strategies in *Ethics and Policy* and 31 occurrences with 5 strategies in *AI Implementation and Usage*. Simultaneously, *Educator Roles* (28 instances, 5 strategies) illustrate the evolving roles of teachers in AI-augmented settings. *Personalization and Optimization* (19 occurrences, 5 strategies) highlight opportunities for tailoring educational experiences, while *Advanced Tools and Technologies* (2 occurrences, 2 strategies) focus on innovative and emerging tools for enhancing teaching and learning.

**Essential Student Skills.** Our literature review identified 14 essential skills for students learning with GenAI, ranked by their frequency of occurrence in the reviewed studies depicted in Table 7. The frequency analysis indicates notable patterns in the literature. *Critical and Analytical Thinking* was the most commonly recognized skill (15), closely followed by *AI Literacy and Awareness* (14) and *Ethical and Responsible AI Practices* (14). These findings underscore the significance of essential competencies that students must possess for comprehending, evaluating, and ethically managing AI-generated content. *Adaptability and Resilience*, *AI Output Evaluation*, and *Problem-Solving and Decision-Making* were each moderately highlighted, occurring 8 times. The practical skills represent essential capabilities for students to navigate GenAI tools, which constantly transform in their operational nature.

Skills such as *Data Privacy and Fairness* and *Prompt Engineering* (6 occurrences each) further underscore the need for understanding AI models deeply and using those insights to write prompts that guide the models to get the desired output while maintaining the ethical implications in generative Artificial intelligence usage. The reviewed literature demonstrates the limited significance of *Programming and Creativity* as well as *Social Skills* and *Questioning and Inquiry Skills* for student interaction with GenAI. This research lays the groundwork for identifying the essential skills that require enhancement, thereby equipping students to adeptly employ GenAI within their academic pursuits.

Table 7. Importance of Student Skills in Literature

| No. | Integration and Curriculum Design | Count |
|---|---|---|
| 1 | Critical and Analytical Thinking | 15 |
| 2 | AI Literacy and Awareness | 14 |
| 3 | Ethical and Responsible AI Practices | 14 |
| 4 | Adaptability and Resilience | 8 |
| 5 | AI Output Evaluation | 8 |
| 6 | Problem-Solving and Decision-Making | 8 |
| 7 | Data Privacy and Fairness | 6 |
| 8 | Prompt Engineering Skills | 6 |
| 9 | Self-Regulation and Reflection | 5 |
| 10 | Bias Awareness and Management | 4 |
| 11 | Creativity and Idea Generation | 2 |
| 12 | Programming and Technical Skills | 2 |
| 13 | Questioning and Inquiry Skills | 2 |
| 14 | Social, Emotional, and Communication Skills | 2 |

## 4.2 Survey Findings

**Background and Familiarity with GenAI.** The survey included 130 participants from diverse academic and cultural backgrounds. The demographic characteristics further illustrated an equitable distribution of gender: 56 female and 61 male respondents, alongside a smaller group of 13 participants identifying as other genders (e.g., non-binary, transgender, gender variant, etc.). Participants represented 20 nationalities, with the majority from Pakistan (47), Germany (32), and India (32), reflecting a global perspective but with a focus on South Asia and Europe. Academically, most respondents were pursuing a Master's degree (87), followed by Bachelor's programs (35) and PhDs (8), indicating strong representation of postgraduate students. In terms of academic disciplines, the largest group belonged to STEM (Science, Technology, Engineering and Mathematic) fields (73), followed by Business and Management (36), with smaller groups from Social Sciences (11), Arts and Humanities (6), and a few specialized fields.

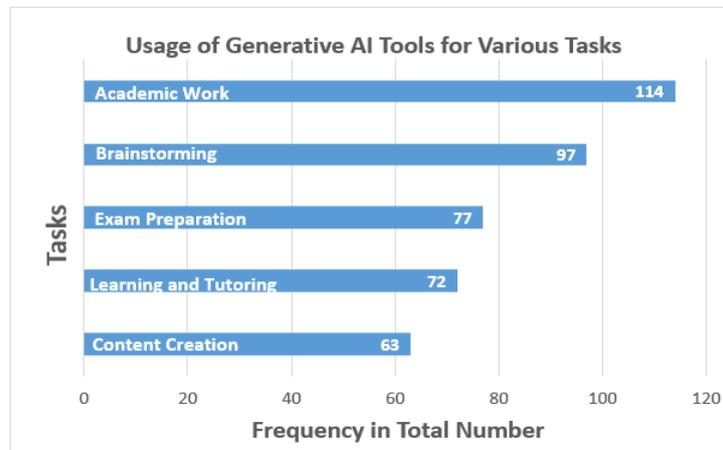

**Fig. 2.** Prior Use of GenAI Tools by Task Type

The survey participants' previous usage of GenAI tools across different tasks is illustrated in Fig. 2. For this question, participating individuals were allowed to choose multiple responses due to the diverse capabilities of these technologies. The majority of participants used GenAI tools for academic purposes, including assignments and thesis research, while the second most common use involved brainstorming and generating ideas, followed by exam preparation and learning/tutoring activities. Content creation, such as blogs and social media, was less frequent but still notable.

**Skills Self-Assessment.** The self-assessment results, as presented in Table 8, highlight varying proficiency levels among participants across eight key skills essential for engaging with GenAI tools. These skills were derived from Table 7, where the original 10 essential skills were streamlined into eight categories to minimize cognitive load for participants during the survey. For instance, *Critical and Analytical Thinking* was combined with *AI Output Evaluation*, while *Ethical and Responsible AI Practices* were merged with *Data Privacy and Fairness* to create more comprehensive skill groupings.

The mean scores in Table 8 indicate that students generally rate themselves as moderately skilled in GenAI-related competencies, with scores ranging from 3.2 to 3.4 on a 5-point Likert scale. Among these, *Understanding of AI Concepts* (Mean = 3.37) and *Evaluating AI Outputs* (Mean = 3.34) are the strongest areas, while *Ethical Decision-Making* (Mean = 3.22) and *Logical Reasoning with AI Tools* (Mean = 3.28) received slightly lower ratings, suggesting a need for further support in these domains. The weighted standard deviation values (ranging from 0.93 to 1.02) show moderate variability in student responses. Skills like *Logical Reasoning* (SD = 1.02) exhibit higher variation, indicating that students have mixed proficiency levels in this area, whereas *Understanding AI Concepts* (SD = 0.94) shows more consistent self-assessments.

Similarly, Effectiveness in Utilizing AI Insights for *Problem-Solving* emerged as a key strength, with 54 participants rating themselves as *Skilled* and 12 *Very Skilled*, highlighting students' ability to apply GenAI in practical scenarios. Moderate profi-ciency was observed in areas like *Evaluating AI-generated outputs* and *Ethical Decision-Making,* where most participants rated themselves as *Moderately Skilled* (52 and 49, respectively) or *Skilled* (41 and 42, respectively). This indicates a foundational understanding but a need for further development in critical thinking and ethical application. In contrast, among the skills assessed, *Confidence in Crafting and Refining AI Prompts* and *Handling Unexpected or Incorrect AI Results* received lower ratings. A large portion of 52 participants identified their *Prompt Engineering* expertise as *Moderate,* but 24 indicated insufficient knowledge ranging from slight to complete absence of skill sets. Similarly, the ability to *Handle Unexpected or Incorrect AI Results* had 33 *Slightly Skilled* and 6 *Not Skilled* responses, reflecting challenges in managing AI errors effectively. The survey data revealed low proficiency in *Self-Regulation* together with Learning Reflection skills since 62 participants showed *Moderate Skills,* but just 9 participants demonstrated strong abilities in these areas.

Overall, the findings demonstrate strong foundational knowledge and problem-solving abilities while revealing a necessity for further training to enhance practical skills, *ethical understanding* and *self-regulation*. The gathered insights will serve as a roadmap for designing curricula which help students perform more efficiently with GenAI systems.

Table 8. Survey results clustered in general topics represented on a 5-point Likert scale (1 Not Skilled, 2 Slightly Skilled, 3 Moderately Skilled, 4 Skilled, 5 Very Skilled)

| | Topic | 1 | 2 | 3 | 4 | 5 | Mean | SD |
|---|---|---|---|---|---|---|---|---|
| Critical thinking and evaluation | Ability to Evaluate AI-Generated Outputs | 5 | 17 | 52 | 41 | 15 | 3.34 | 0.97 |
| | Confidence in Logical Reasoning with AI Tools | 6 | 22 | 46 | 42 | 14 | 3.28 | 1.02 |
| AI Literacy | Understanding of AI concepts | 5 | 15 | 49 | 49 | 12 | 3.37 | 0.94 |
| | Familiarity with Generative AI in Education | 3 | 24 | 44 | 46 | 13 | 3.32 | 0.96 |
| Ethical AI and data Management | Ethical Decision-Making with AI Tools | 7 | 21 | 49 | 42 | 11 | 3.22 | 0.99 |
| | Awareness of Ethical and Fairness Issues in AI | 8 | 30 | 52 | 26 | 14 | 3.06 | 1.05 |
| | Responsibility in Data Management and Privacy with AI | 8 | 26 | 50 | 34 | 12 | 3.12 | 1.03 |
| Adaptability and resilience | Balancing AI Use with Personal Effort | 6 | 30 | 38 | 43 | 13 | 3.21 | 1.05 |
| | Handling Unexpected or Incorrect AI Results | 6 | 33 | 46 | 27 | 18 | 3.14 | 1.09 |
| Problem-solving | Effectiveness in Utilizing AI Insights for Problem-Solving | 6 | 17 | 41 | 54 | 12 | 3.38 | 0.98 |
| Prompt engineering skills | Confidence in Crafting and Refining AI Prompts | 7 | 17 | 52 | 37 | 17 | 3.31 | 1.03 |
| Self-regulation | Learning Reflection and Decision Ownership with AI | 6 | 18 | 62 | 35 | 9 | 3.18 | 0.92 |
| Bias awareness | Ability to Identify Biases in AI Outputs | 5 | 22 | 50 | 43 | 10 | 3.24 | 0.95 |

**New Skills and Support Needs.** This section presents an analysis of students' perspectives on essential skill development, institutional support, and teaching strategies for GenAI using three Pareto charts. The Pareto principle (80/20 Rule) is applied to identify the "vital few" elements that contribute the most to students' learning needs. This Principle, developed by Vilfredo Pareto, state that 80% of results are determined by 20% of activities [34].

Since the original weighted averages for responses were closely grouped, the data was normalized to a 0–1 scale using the following formula:

$$X_{norm} = \frac{(X - X_{min})}{(X_{max} - X_{min})}$$

[27]

Where $X$ is the original value, $X_{min}$ is the minimum value in the dataset, $X_{max}$ is the maximum value in the dataset, $X_{norm}$ is the normalized value of $X$.

**Key Findings on Student Needs for Skill Growth.** This Pareto chart analyses the areas where students require support to effectively engage with GenAI tools. In our survey data, students were asked to rate the level of support they need for developing specific GenAI-related skills. To ensure accurate visualization, a weighted average of responses under each skill was calculated and then normalized, allowing us to better highlight the most significant areas where support is needed.

Given the length of the skill names, they are represented using legends rather than direct labels on the x-axis. The blue and grey bars depict the normalized weighted averages of the responses where students need support in using GenAI, arranged in descending order. The orange line illustrates their cumulative contribution, following the Pareto principle (80/20 rule), which suggests that a small set of key skills accounts for the majority of student challenges.

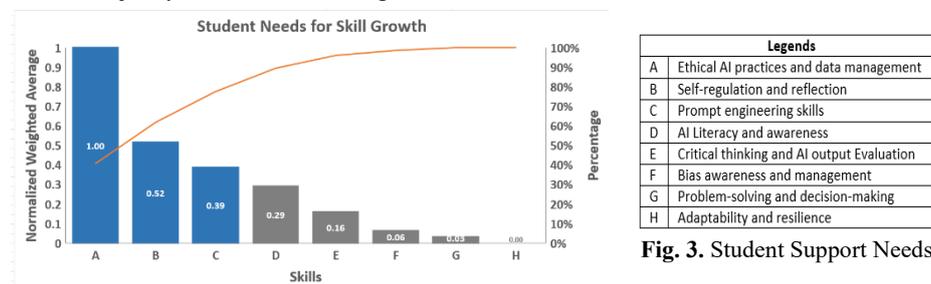

**Fig. 3.** Student Support Needs

Overall, the blue-highlighted bars indicate high-support areas, with *Ethical AI practices and data management*, *Self-regulation and reflection*, and *Prompt Engineering* emerging as the top three skill gaps. These three skills alone account for the majority of the cumulative importance for almost 80%, meaning that addressing them would significantly improve students' overall engagement with GenAI tools. This highlights the need for structured training programs focused on AI ethics, self-directed learning, and effective prompt crafting. While other skills like *AI Literacy*, *Bias awareness and management*, and *Critical thinking and AI output Evaluation* also show a need for development, they are relatively less critical compared to the highest-ranked skills.

**Key Findings on Institutional Support for Skill Development.** This Pareto chart analyzes students' preferences regarding institutional support for developing GenAI-related skills. Based on survey data, students rated the importance of various forms of institutional support, allowing us to identify which strategies should be prioritized. To ensure better visualization, the weighted average for each form of institutional support was calculated and normalized, helping to highlight the most preferred form of institutional support by students.

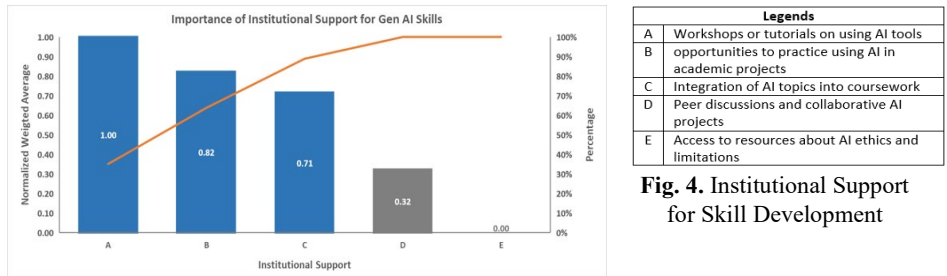

**Fig. 4.** Institutional Support for Skill Development

Overall, the blue-highlighted bars indicate the most preferred institutional support types by students. These include Workshops or tutorials on using AI tools, project-based learning opportunities, and Integration of AI topics into coursework. These three forms of support alone account for the majority of the cumulative importance for almost 80%. Prioritizing these "vital few" support types would cater to most of the students' skill development needs, ensuring that AI education is both practical and accessible. Beyond workshops and project-based learning, Peer discussions and collaborative projects involving AI also play a significant role in student skill enhancement. While these areas were rated moderately important, integrating AI literacy within coursework and fostering collaborative discussions can create a more well-rounded approach to AI education

**Key Findings on Adjusting Teaching Strategies in Response to GenAI.** This Pareto chart analyzes students' preferences regarding the importance of various teaching strategies for integrating GenAI into education. Based on survey responses, students rated six key lecturer strategies, which were identified from the literature review as the most frequently cited approaches for GenAI-driven teaching adaptation.

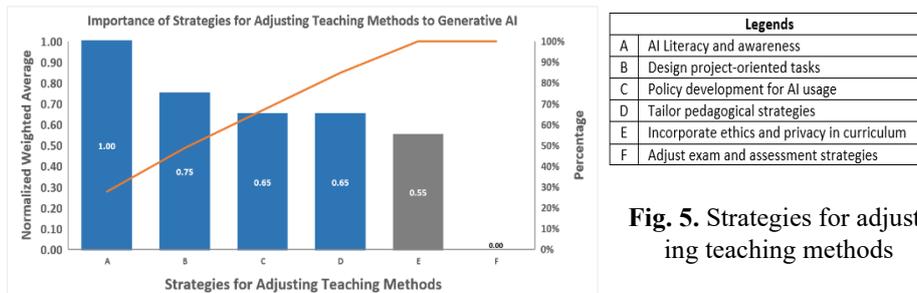

**Fig. 5.** Strategies for adjusting teaching methods

Overall, the blue-highlighted bars indicate the most critical teaching strategies, with A*I Literacy and awareness and Design project-oriented tasks* emerging as the top priorities. These strategies alone account for a majority of cumulative importance, suggesting that focusing on them would maximize the impact of GenAI integration in education. AI Literacy is essential for providing students with a structured understanding of AI technologies, while *Project-Based Learning* emphasizes hands-on application, aligning with student preferences for practical AI engagement. Other strategies, such as *Policy Development for AI Usage*, *Tailoring Pedagogical Ap-* *proaches*, *Incorporating Ethics and Privacy*, and *Adjusting Exam and Assessment Strategies*, were also considered important but contributed less to the cumulative significance. However, they should not be overlooked, as they provide a complementary role in ensuring responsible GenAI adoption, ethical awareness, and adaptable teaching methods.

## 5 Discussion

The research examined fundamental competencies students must possess for effectively engaging with GenAI tools and also compiled various strategies for educators in six broad categories to integrate GenAI into their teaching methods. The study also assessed students' self-perceived proficiency in GenAI-related skills, their support needs, and their perspectives on institutional support through the survey form. The findings, obtained from a literature review and survey analysis, provide significant insights into how educational institutions may enhance student preparedness for AI-driven learning environments.

In the literature review, 14 essential competencies for GenAI engagement were identified. Among them, *Critical and Analytical Thinking* together with *AI Literacy and Awareness* and *Ethical and Responsible AI Practices* appeared most often. Students evaluated most of these skills through an online survey at a moderate level. However, gaps were identified in *prompt engineering*, *bias awareness*, and *handling unexpected AI outputs*. The Pareto analysis of support needs further highlighted these gaps with the students who required the most assistance in three primary domains: *ethical AI practices, self-regulation, and prompt engineering skills* (see Fig. 3). These three areas alone accounted for over 80 percent of cumulative importance, indicating that focusing on these competencies would address the majority of student challenges.

In both literature and survey findings, the importance of institutional support for GenAI adoption is emphasized. The study analyzed educator strategies, which were compiled into six distinct categories. The survey revealed students' top preferences for institutional support through Pareto analysis. The strategy of integrating GenAI tools into the educational curriculum received the highest recognition from both scholarly texts and survey participants, with thirteen instances in the literature review. This reflects a strong consensus that embedding GenAI tools into coursework and assessments is crucial for student success. Students ranked *AI Literacy and Awareness* as their first choice for academic priority since they require structured instruction about AI fundamentals. The literature demonstrated *Adjusting Exam and Assessment Strategies* as a frequently cited strategy (12 occurrences). The survey revealed *project-oriented tasks* to be the most important educational priority according to students with almost 50 percent cumulative importance weight. This suggests a shift toward hands-on, real-world applications of GenAI tools over traditional exam reforms, reinforcing the need for experiential learning approaches.

The literature reveals the importance of *policy development* (11 occurrences) for GenAI, which students also view as essential institutional support (see Fig. 5). Interestingly, the literature focused heavily on including ethics and privacy education in academic lessons (10 occurrences), but students assigned lower importance to these topics. Students demonstrate a preference for learning practical GenAI implementa-tion skills over theoretical considerations in ethical cases. Although they differ in some aspects, the two views support the need to develop ethical policies for GenAI to ensure responsible use. Students also highlighted the value of *peer discussions and collaborative GenAI projects* (30% cumulative importance), which aligns with the literature's emphasis on the educator's role as a facilitator rather than a sole content provider.

The survey results indicate that educational institutions need to make GenAI practical training their main priority alongside teaching core concepts in their curriculum. Various educational approaches that include *hands-on workshops,* along with *project-based learning* and AI literacy programs, create paths to address the skill deficiencies noted by survey respondents. The development of established policies stands as essential because students requested explicit institutional guidelines about responsible GenAI use. Technical training for future education professionals must receive ethical and sustainable GenAI integration support through policies that address both AI ethics standards and academic integrity guidelines, as well as responsible AI usage principles.

The difference between students' preferences and academic literature recommendations regarding ethics versus hands-on learning shows that educational institutions need to find a balance between theoretical approaches and practical applications for delivering effective GenAI critical skills.

Research data showed that 87% of students use GenAI tools for their academic work involving assignments, thesis research and project assignments. A large percentage of students using GenAI tools demonstrates an urgent need for institutional guidelines that promote both GenAI usage responsibility and ensuring academic integrity standards. Additionally, 59% of participants adopt GenAI technology for exam preparation, while 75% use GenAI for brainstorming. This demonstrates how GenAI supports both innovative thinking and independent study techniques. Institutions can leverage this trend by integrating AI-driven ideation tools or the use of chatbots into curricula and promoting GenAI as a study aid in exam preparation workshops.

## 6      Conclusion

This research paper explored necessary student competencies working with GenAI tools while presenting effective strategies to facilitate educators in GenAI integration. The data from the literature analysis and survey findings highlights gaps in GenAI-related competencies for students while offering direction for lecturers to deal with the evolving landscape of AI-driven learning. Although all the skills identified in the literature and the strategies proposed for educators hold immense importance for effectively integrating GenAI into education, this study highlights specific areas that emphasize the need for immediate attention. We found that AI literacy, together with critical thinking abilities and ethical AI fundamentals, are essential base skills students need to learn. Our research shows shortcomings in prompt engineering practice and AI bias recognition together with handling AI-generated content. The Pareto analysis of student responses revealed that focusing on *ethical AI practices, self-*

...

*regulation*, and *practical AI applications* would address the majority of students' challenges.

Both the literature review and survey findings indicate that AI education must be included in project-based learning methods. While educators and policymakers emphasize the need for ethical considerations, students prioritize hands-on, experiential learning with GenAI tools. This divergence illustrates the essential requirement for educational institutions to strike a balance between theoretical paradigms and practical applications, thereby ensuring that students cultivate both the technical expertise and ethical competencies. The extensive adoption of GenAI tools by students throughout academic work, including assignments and thesis research, reveals an immediate requirement for institutional guidance that supports academic standards and ethical AI usage.

However, this study has some limitations. The geographical generalization of the research results is limited, as well as sample size was rather constrained. Augmenting the number of participants could improve the robustness and generalizability of the findings. Another notable limitation of this study is the potential influence of the Dunning-Kruger Effect [36] on self-assessment results. This cognitive bias suggests that individuals with low competence in a given domain often overestimate their abilities, while those with higher competence tend to underestimate their expertise. Since this study relied on self-reported proficiency ratings, participants may have unintentionally misjudged their actual skill level.

Building on this study's findings, several areas warrant further investigation to enhance the integration of GenAI in education. Future studies should incorporate objective skill-testing methods, such as AI-based problem-solving exercises or standardized assessments, to compare perceived versus actual competencies. Additionally, longitudinal studies could track students' AI skill development over time, offering insights into how GenAI proficiency evolves with continued exposure and training. Another important direction for future studies should explore cross-cultural differences in AI adoption, skill gaps, and institutional support, as GenAI integration strategies may vary significantly across different educational systems.

In conclusion, the successful integration of GenAI into education requires a multifaceted approach, while the literature review provides a solid foundation for identifying essential skills and teaching strategies related to GenAI in education. However, given the fast-paced evolution of GenAI, new research is constantly emerging alongside the development of innovative GenAI tools, such as the recent launch of DeepSeek- R1. The foundation laid by this study offers room for future investigations to combine present-day AI developments with new skill development and teaching methods that will appear because of AI technological progress.

**Acknowledgments** The research was supported by the German Federal Ministry of Education and Research (BMBF) in the Programme Künstliche Intelligenz in der Hochschulbildung (Grant No. 16DHBKI010) as well as the Programme Sachsen-Anhalt WISSENSCHAFT Gleichstellung, Qualifikation, Nachwuchs aus dem Europäischen Sozialfonds Plus (Grant No. ZS/2023/11/181808). We thank all students who voluntarily participated in the online survey. Furthermore, we would like to acknowledge the insightful comments and constructive suggestions from Frieder Stolzenburg.